\documentclass[12pt]{article}
\title{Improvement of perturbation theory in QCD for
$e^+e^-\to hadrons$ and the problem of $\alpha_s$ freezing.}
\author{B.V.Geshkenbein\thanks{e-mail:geshken@vitep5.itep.ru} and
B.L.Ioffe\thanks{e-mail:ioffe@vitep5.itep.ru}\\
\\
Institute of Theoretical
and Experimental Physics
\\
117218, Moscow, Russia}
\date{}

\begin{document}

\maketitle

\newcommand{\be}{\begin{equation}}
\newcommand{\ee}{\end{equation}}

\def\la{\mathrel{\mathpalette\fun <}}
\def\ga{\mathrel{\mathpalette\fun >}}
\def\fun#1#2{\lower3.6pt\vbox{\baselineskip0pt\lineskip.9pt
\ialign{$\mathsurround=0pt#1\hfil##\hfil$\crcr#2\crcr\sim\crcr}}}

\begin{abstract}

We develope the method of improvement of perturbative theory in QCD, applied
to any polarization operator. The case of polarization operator $\Pi(q^2)$,
corresponding to the process $e^+e^-\to hadrons$  is considered in details.
Using the analytical properties of $\Pi(q^2)$  and perturbative expansion of
$\Pi(q^2)$ at $q^2< 0$, $Im \Pi(q^2)$ at $q^2 >0$ is determined in such a
way, that the infared pole is eliminated. The convergence of perturbative
series for $R(q^2)=\sigma(e^+e^-\to hadrons)/(e^+e^-\to \mu^+\mu^-)$ is
improved. After substitution of $R(q^2)$ into dispersion relation the
improved Adler function $D(q^2)$ with no infrared pole and frozen
$\alpha_s(q^2)$   has been obtained. A good agreement with experiment has
been achieved.

\end{abstract}

\hspace{3mm} PACS: 12.38. Cy

\vspace{7mm}

It is well known for years, that perturbative calculations of amplitudes
in field theories are legitimate, if virtualities of all external legs
$k^2_i$ are negative, $k^2_i < 0$ . In this case polarization operators
and vertex functions (two- and three-point functions) are off mass
shell and have no singularities. However, in order to get physical
predictions, it is necessary to go to $k^2_i > 0$ (at least for some of
them). This can be achieved by analytical continuation using  known
analytical properties of amplitudes. A typical example is
$e^+e^-$-annihilation into hadrons in QCD. The total cross section
$\sigma(e^+e- \to hadrons)$ is proportional to imaginary part of the
polarization operator $\Pi_{\mu \nu}(q^2)$ of virtual photon at $q^2 >
0$. At large enough $q^2 < 0 ~~ \Pi_{\mu \nu}(q^2)$ can be calculated
perturbatively in QCD in terms of the expansion over the running
coupling constant $\alpha_s(q^2) = 4 \pi/\beta_0 ln(-q^2/\Lambda^2$).
The tensor structure of $\Pi_{\mu \nu}(q)$ is

\be
\Pi_{\mu \nu}(q) = (q_{\mu} q_{\nu} - q^2 \delta_{\mu \nu}) \Pi(q^2)
\label{1}
\ee
and $\Pi(q^2)$ is an analytical function of $q^2$ in the whole $q^2$
complex plane with a cut along positive $q^2$ semiaxes. Analytical
continuation from $q^2 < 0$ to $q^2 > 0$ results in substitution of
$ln(q^2/\Lambda^2) \to ln (Q^2/\Lambda^2) - i \pi, ~~ Q^2 = \mid q^2
\mid$. Since small $\alpha_s$ corresponds to large
$ln(Q^2/\Lambda^2)$, the standard procedure (see, e.g. \cite{1}) is to
consider $ln(Q^2/\Lambda^2)$ as large comparing with $\pi$ and to
perform the expansion in $ \delta = \pi/ln(Q^2/\Lambda^2$). However,
practically it is not a good expansion parameter. At typical scale of
$e^+e^-$-annihilation, $Q^2 \sim 10~ GeV^2$ and $\Lambda \sim 300 - 400
~MeV, ~~~ ln(Q^2/\Lambda^2) \approx 4 - 5$ and $\delta \approx 0.7$.

In this paper we present the systematic method of improvement of
perturbation theory in QCD, which is free from this drawback. Besides
$e^+e^-$ annihilation, this method can be applied to any polarization
operators, for example, to those used in the QCD sum rule approach. The
idea of the method has been suggested by Radyushkin \cite{2} and considered
also by Pivovarov \cite{3}, but many important features and consequences of
this method were not touched in \cite{2},\cite{3} (particularely, eq.
(\ref{10}) below was not obtained and analysed).
Consider the Adler function

\be
D(q^2) = -q^2~ \frac{d \Pi(q^2)}{d q^2} = -q^2~ \int\limits^{\infty} _0~
\frac{R(s) ds}{(s - q^2)^2}
\label{2}
\ee
where $R(s) = \sigma(e^+e^- \to hadrons)/\sigma(e^+e^- \to \mu^+
\mu^-)$. In the parton model $R(s) = R_p = 3 \sum\limits_q~ e^2_q$,
where $e_q$ is the charge of the quark of flavour $q$. It is convenient
to write:
\be
D(q^2) = R_p(1 + d(q^2)), ~~~ R(q^2) = R_p(1 + r(q^2)), ~~ \Pi(q^2) =
R_p(1 + p(q^2))
\label{3}
\ee
From (\ref{3}), there follows the equation:
\be
d(q^2) = -q^2 \frac{dp(q^2)}{d q^2}
\label{4}
\ee
The solution of (\ref{4}) is
\be
p(q^2) - p(\mu^2) = - \int\limits^{q^2}_{\mu^2}~ \frac{ds}{s} d(s)
\label{5}
\ee
$r(q^2)$ is proportional to discontinuity of $p(q^2)$ at $q^2 > 0$:
\be
r(q^2) = \frac{1}{\pi} Im p(q^2) = \frac{1}{2 \pi i} \Biggl [ p(q^2 + i
\varepsilon) - p(q^2 - i \varepsilon) \Biggr ]
\label{6}
\ee
At negative $q^2 < 0, ~ q^2 = -Q^2$ the perturbative expansion of
$d(Q^2)$ in $\overline{MS}$ renormalization scheme is known up to the
third order \cite{4},\cite{5}:
$$
d(Q^2) = a(1 + d_1 a + d_2 a^2), ~~~~ d_1 = 1.986 - 0.115 N_f
$$
\be
d_2 = 18.244 - 4.216 N_f + 0.086 N^2_f
\label{7}
\ee
where $a(Q^2) = \alpha_s(Q^2)/\pi, ~ N_f$ is the number of flavours and
the small gluon-gluon scattering terms are omitted.  With the same
accuracy, three loops expression for $\alpha_s(Q^2)$ in $\overline{MS}$
is given by \cite{6,7}
\be
a(Q^2) = \frac{4}{\beta_0 L} ~ \left\{ 1 - \frac{2 \beta_1}{\beta^2_0}~
\frac{ln L}{L} + \frac{4 \beta^2_1}{\beta^4_0 L^2} \Biggl [ (ln L -
\frac{1}{2})^2 +
\frac{\beta_2 \beta_0}{8 \beta^2_1}- \frac{5}{4}\Biggr ] \right \}
\label{8}
\ee
where $L = ln(Q^2/\Lambda^2)$
and
\be
\beta_0 = 11 - \frac{2}{3} N_f, ~~ \beta_1 = 51 - \frac{19}{3} N_f, ~~
\beta_2 = 2857 - \frac{5033}{9} N_f + \frac{325}{27} N^2_f
\label{9}
\ee
Substitution of (\ref{7}), (\ref{8}) into (\ref{5}) and (\ref{6}) leads
to perturbative corrections up to the third order in the physically
measurable quantity $r(q^2)$.By taking the discontinuity, any
dependence on the normalization point $\mu^2$ is eliminated. It should
be stressed, that in such calculation no expansion in
$\pi/ln(Q^2/\Lambda^2)$ is performed, the only assumptions used(in
fact, they are not assumptions, but theorems), are the analytical
properties of $\Pi(q^2)$ . The result is $(q^2 \geq 0)$:
$$
r(q^2) = \frac{4}{\pi \beta_0} ~\kappa - 8 \frac{\beta_1}{\beta^3_0}
\frac{1}{\pi^2
\tau^2} \Biggl [ ln(\pi \tau) + 1 - \kappa t \Biggr ] + \Biggl (
\frac{4}{\beta_0} \Biggr )^2 \frac{d_1}{\pi^2}~ \frac{1}{1 + t^2} +
$$
$$
+ 16 \frac{\beta^2_1}{\beta^5_0}~ \frac{1}{\pi^3 \tau^4}
\left \{ \Biggl (\frac{\beta_2 \beta_0}{8 \beta^2_1}  - 1 \Biggr ) t + \kappa
(1 - t^2) ln (\pi \tau) + t [ln^2 (\pi \tau) - \kappa^2 ] \right \} -
$$
\be
-4d_1 \Biggl (\frac{4}{\beta^2_0} \Biggr )^2~ \frac{\beta_1}{\pi^3
\tau^4} \left \{ t \Biggl [ln (\pi \tau) + \frac{1}{2}\Biggr ] +
\frac{1}{2}\kappa (1 - t^2) \right \} + d_2 \Biggl ( \frac{4}{\beta_0}
\Biggr )^3~ \frac{t}{\pi^3 \tau^4}
\label{10}
\ee
where
\be
t = \frac{1}{\pi} ln \frac{q^2}{\Lambda^2}, ~~ \tau = (1 + t^2)^{1/2},
~~ \kappa = \frac{\pi}{2} - arctg ~t
\label{11}
\ee
Essential features of (\ref{10}) are: 1) unlike standard perturbation
expansion $r(q^2)$ has no infra-red poles at $q^2 = \Lambda^2$ and
tends to finite limit at $q^2 \to 0,~~ r(0) = 4/\beta_0 \approx 0.414$,
corresponding to $\alpha_s(0) = 4 \pi/\beta_0 \approx 1.3$, all higher order
terms vanish; 2) the convergence of perturbation series in (\ref{10}) is
much better, than of the standard ones: the ratio of the second order term
to the first order one is smaller than 0.2 everywhere and the ratio of the
third order term to the second one is smaller than 0.1, while in the
standard approach the ratio of the third order term to the second order one
exceeds 0.5 below $q^2 = 10~ GeV^2$ and is larger than 1 at $q^2 \la 1
~GeV^2$;~~ 3) the $\alpha_s$ corrections, given by (\ref{10}), are
remarkably smaller in the low energy domain, $q^2 \la 5~ GeV^2$, than the
standard ones, e.g., at $q^2 = 1 GeV^2$ $r$(eq.10)$/r_{stand} = 0.72$.

In order to get $r(q^2)$ at some lower values of $q^2$, starting from
the value of $\alpha_s$ at $Z$-boson mass, $\alpha_s(m_z)$, which is
rather well known now, it is necessary to use renormalization group
equations (\ref{8}) and perform matching at the thresholds of new
flavours production $b$ and $c$ quarks and, if we would like to go to
$q^2 < 1~ GeV^2$, even at $s$-quark threshold. This matching procedure
introduces some uncertainty. The matching may be performed at $2 m_q$
(or, what is practically equivalent,  at $m_{\Upsilon}$ and
$m_{J/\psi}$) or at $m_q$. The former seems to be preferable in the case of
$e^+e^-$-annihilation for the evident reasons. There are some arguments in
the favour of the latter choice, based on minimal sensitivity of the results
to the matching point value \cite{8}. In the standard approach the results
are rather sensitive to the choice of matching points -- the ratio of $r$'s
in the two above mentioned cases is about $1.10 - 1.15$ at $q^2 \la 5~
GeV^2$.
We calculated the $q^2$  dependence of $r(q^2)$  starting from the point
$q^2=m^2_z$   and going down. The value $r(m^2_z)$  was found from the
requirement, that $\alpha_s(m^2_z)$ determined in the standard way is equal
$\alpha_s(m^2_z)=0.119\pm 0.002$ \cite{9}. Then
$\Lambda_5=230^{+27}_{-27}~MeV$ was found. In the evolution down to lower
energies the matching  of $r(q^2)$ at the masses of $\Upsilon,~J/\psi$  and
$\varphi$  was performed resulting in $\Lambda_4=335^{+35}_{-33}~MeV$,
$\Lambda_3=414^{+41}_{-37}~MeV$, $\Lambda_2=490^{+51} _{-47}~MeV$. It was
found particularly, that $\pi r(1~GeV^2)=0.41$. If, instead of matching at
quarkonium masses the matching at $m_q$  would be done, the values of
$r(q^2)$  would be only 4\% smaller below $q^2=10~GeV^2$  practically
independent of $Q^2$. Therefore, in this aspect this method has also some
advantages. The performed comparison with experiment demonstrated a good
agreement, starting from $\sqrt{q^2}=0.7~GeV$, i.e. in much broader
interval, than in recent paper \cite{10}.

After substituting $r(q^2)$, given by (\ref{10}) into dispersion
relation (\ref{2}) we get back Adler function $D(q^2).$ Obtained in this way
improved $D(q^2)_{impr}$  has correct analytical properties and no
unphysical singularities. Therefore, the adopted here procedure  has a
serious advantage: the required analytical properties of Adler function are
restored. If $D(q^2)_{impr}$  is represented in terms of improved effective
QCD coupling constant $\alpha_s(q^2)_{impr}$, this would mean, that
$\alpha_s(q^2)_{impr}$  has no infrared pole and is frozen at $q^2\to 0$. In
this aspect our approach has some resemblance to the approach by Shirkov and
Solovtsov \cite{11} (see also \cite{12}), where the condition of analyticity
of $\alpha_s(q^2)$ in the cutted $q^2$  complex plane was imposed. The
difference is, that we exploit the analyticity of polarization operator
$\Pi(q^2)$, which is a strict result in field theory and do not use any
hypothesis about analyticity of $\alpha_s(q^2)$.

The method to define the improved QCD coupling constant
$\alpha_s(q^2)_{impr}$  through $D(q^2)_{impr}$   looks  very promisable.
For this goal, perhaps, the most suitable is to use not $\overline{MS}$, but
the Brodsky, Lepage, Mackenzie renormalization scheme \cite{13}, where
$e^+e^-$-annihilation is considered as a basing process for definition of
$\alpha_s(q^2)$  with no higher order $\alpha_s$ corrections. This problem
requires further investigation.

The presented above method can be applied to the treatment of perturbative
corrections to any polarization operators and, therefore may result to
improvement of QCD sum rule approach.

This work was supported in part REBR grant 97-02-16131.



\vspace{5mm}

\begin{thebibliography}{99}
\bibitem{1} J.D.Bjorken, SLAC-PUB-5103 (1989).
\bibitem{2} A.V.Radyshkin, JINR preprint E2-82-159 (unpublished), JINR
Rapid Communications, 4[78]-96, 9 (1996).
\bibitem{3}  A.A.Pivovarov, Nuovo Cim. {\bf 105A}, 813 (1992).
\bibitem{4}   K.G.Chetyrkin, A.L.Kataev and F.V.Tkachev, Nucl.Phys., 
{\bf B174}, 345 (1980).  
\bibitem{5}  A.L.Kataev and V.V.Starshenko, 
Mod.Phys.Lett {\bf A10}, 235 (1995).  
\bibitem{6}  O.V.Tarasov, 
A.A.Vladimirov and A.Yu.Zharkov, Phys.Lett. {\bf B93}, 429 (1980).  
\bibitem{7}  S.A.Larin and J.A.M.Vermaseren, Phys.Lett. {\bf B303}, 304
(1993).
\bibitem{8}  W.Bernreuther, Ann.Phys. {\bf 151}, 127 (1983).
\bibitem{9}  Review of Particle Physics, Eur.Phys.J {\bf C3}, 1
(1998).
\bibitem{10} S.Eidelman,
F.Jegerlehner, A.L.Kataev and O.Veretin, preprint DESY-98-206,
hep-ph/9812521.
\bibitem{11}  D.V.Shirkov and I.L.Solovtsov, Phys.Rev.Lett. {\bf 79}, 1204
(1997).
\bibitem{12} E.Gardi, G.Grunberg and M.Karliner, JHEP, {\bf 9807}, 007
(1982).
\bibitem{13}  S.Brodsky, G.P.Lepage and P.B.Mackenzie, Phys.Rev.
{\bf D28}, 228 (1983).
\end{thebibliography}
\end{document}